# Aggregation of colloidal nanoparticles in polymer matrices

Soft Matter, Emerging Area
*Julian Oberdisse*

*Laboratoire des Colloïdes, Verres et Nanomatériaux (LCVN)*
*Université Montpellier II*
*34095 Montpellier, France*

## *revised version*

28[th] of September 2005

**Abstract**

Colloidal nanoparticles may possess many functional properties, whose nature may be electronic, chemical, biological, mechanical, etc. It is often advantageous to incorporate them into a matrix material, e.g. a polymer solution or melt, or an elastomer, in order to obtain a 'nanomaterial' with additional properties brought in by the filler particles. One of the basic but nonetheless crucial properties is the mechanical strength of such polymer nanocomposites, whose rheological (or mechanical) properties are usually better than those of the pure matrix. The precise origin of this mechanical reinforcement effect, however, remains unclear.

In this context, some recent studies of the structure and mechanical properties of a special type of nanocomposites are reviewed here. In silica-latex systems, a latex film with silica inclusions is formed from a colloidal solution of both components. During drying of the solution, the formation of silica domains can be controlled via the physico-chemical properties of the solution. Well-defined silica aggregates embedded in a polymer matrix can be generated, and the mechanical properties of the resulting nanocomposite have been shown to be directly correlated to the average structure. We believe that the fine-tuning of the structure of the filler phase opens new perspectives for systematic studies of the reinforcement effect, e.g. by modifying filler-polymer interfacial properties at fixed structure, or by generating original structures.

**Figures: 10**

**Introduction**

Viscoelastic material is used for very different industrial applications, from coatings to car tyre engineering, due to the wide range of accessible properties [1,2]. These materials, which are usually based on polymer, need to meet a certain number of criteria corresponding to the desired application, and moreover be economically efficient. Among the more physical requirements, mechanical[1] and optical properties come immediately to one's mind, followed by more specialised features. Concerning the mechanical properties, mankind has found out quite early that those of natural rubber, *hevea brasiliensis*, could be improved – or reinforced - by adding small filler particles [3]. Since then many different filler particles have been employed, like the colloidal silica we report on in this review, or carbon black [4]. In spite of considerable research efforts, leading to great successes of both empirical descriptions and fundamental theories based on microscopic features (like the fractal geometry of aggregates, or tube models for the surrounding polymer), the reinforcement effect still resists a comprehensive description. Different aspects to reinforcement, however, have been identified in specific systems. In this introduction, we will first review the most important contributions, before we focus on a particular model system allowing the isolation of a dominant one, filler structure and aggregation.

In the beginning, only rather big fillers were available. With micron-sized or bigger particles, optical properties like transparency were poor because of the strong scattering and absorption of light by large heterogeneities. As smaller and smaller filler were used, both optical and mechanical properties improved. The latter is mainly due to the increase in specific surface, at least if particles are well dispersed in the matrix. Indeed, typical colloidal particles used as filler are in the 5 - 100 nm range, and the corresponding specific surface can be as high as a few hundred $m^2$/g. Polymer chains can adsorb onto the surface, thereby adding crosslinks of high functionality to the matrix.

Another effect contributing to the reinforcement is induced by the great difference in modulus of the filler and the polymer. As a rule of thumb, a polymer matrix has a (Young's) modulus of about 1 MPa, whereas silica is about 30'000 times stiffer. Therefore, filler particles are virtually not deformed, and most of the macroscopic deformation is stored in disproportionally stretched polymer chains [5]. The latter carry thus a higher stress, which leads to a reinforced modulus of the composite.

---

[1] For the sake of simplicity, we will speak of mechanical properties for both elastomers (solids) and polymer melts (liquids).

Recently, other subtle effects have been shown to exist, like the possible formation of a glassy layer (modulus ≈ 1 GPa) of chains close to rigid surfaces, which increases the effective fraction of hard filler [6-8]. These examples demonstrate to what extent the exact nature of reinforcement is far from being understood, mainly because many different effects may be present and interfere, depending on the detailed chemistry, the glass transition temperature of the matrix, the surface interactions, etc.

Up to this point we have neglected the structure of the filler particles within the polymer matrix. In this review, it will be shown that the state of dispersion and aggregation has a strong influence on the mechanical properties. In particular, we will concentrate on work with rather monodisperse spherical particles, but it is worth mentioning that a strong effort is undertaken by the international community with high-aspect ratio fillers, like clay platelets, fibers or carbon nanotubes, see ref. [9] and references therein.

In a polymer matrix, the state of dispersion of spherical particles can vary from highly dispersed to totally aggregated, depending on the thermodynamics of the system and the kinetics of sample preparation. The clustering of filler particles is in general favoured by strong attractive Van der Waals forces always present between colloidal particles at short distances, whereas steric and long-range electrostatic forces may stabilise individual beads. If aggregation occurs, the internal structure of aggregates ranges from close-packed clusters to tenuous fractals, depending on the system and preparation, with far-reaching consequences for the mechanical response of such samples. In certain cases, a hard filler network spanning the whole sample leads to a tremendous increase in modulus [10]. In other cases, collisions and breaking-up of aggregates introduce additional dissipative mechanisms. When probing the mechanical response of a composite, the structural contribution to reinforcement may therefore dominate all other contributions.

Theoretical studies have been performed in order to correlate filler structure with the mechanical properties of the composite. Analytical mean field theories usually start with the hydrodynamic effect of filler particles, i.e. Einstein's equation for the viscosity of a dilute colloidal solution [11], which has been shown to apply for the modulus of a composite by Smallwood in the 1940's [12]. Further developments include empirical divergences of the modulus at a given volume fraction in order to account for the formation of a continuous hard filler network [13]. Witten et al have applied statistical mechanics in order to describe the break-up of large fractal aggregates [14]. The hydrodynamic reinforcement effect of fractal aggregates has been studied by Huber and Vilgis [15], and recent progress has been reviewed by Heinrich and Klüppel [16].

Probably the most difficult problem for testing such theories is that it is impossible in most experimental systems to disentangle the different contributions, e.g. structural and interfacial, to the reinforcement effect. It is one of the key advantages of the silica-latex model system presented here that the filler structure can be varied over a large range, while the interface is kept almost unchanged. The aggregate size and distribution in space can be measured in the bulk, moreover by a single technique - Small Angle Scattering - which avoids calibration problems, and it can be applied even under deformation. Although it can not be excluded that some interfacial effects persist, these can be shown to be of second order. This allows a direct comparison to theoretical predictions for the reinforcement effect of aggregates of spherical particles incorporated in a polymer matrix.

**Sample preparation**

Unfortunately, pouring a fine powder of nanoparticles onto a polymer melt is not the most efficient way to disperse the particles. To start with, many commonly used polymers are hydrophobic, while for instance colloidal silica is hydrophilic, and even if one hydrophobizes the filler, the high viscosity of the melt impedes the dispersion. Different procedures have been developed to incorporate the filler in the matrix [17]: mechanical mixing (milling), sol-gel techniques, by e.g. introducing the filler before polymerisation, or in situ generation of the filler. The latex route reviewed here relies on colloidal mixtures which are dried [18-21]. Its principle is illustrated in Fig. 1. An aqueous colloidal solution of a hydrophobic latex carrying a shell of ionisable groups of methacrylic acid for electrostatic stabilisation is mixed with a colloidal solution of silica particles. Due to the methacrylic acid, the latex particles carry a negative electric charge at high pH, as do the silica particles due to the silanol groups. The mixture is degassed and dried above the film formation temperature which is close to the glass transition temperature of the polymer. The polymer particles are then liquid drops which coalesce as the water is removed to form a continuous polymer film. Due to the presence of the ionisable groups, and to the silanol groups on the silica surface which have the same function of charge stabilisation, the structure of these solutions is pH- and electrolyte responsive. We will see here that the final structure of the silica particles embedded in the polymer matrix can be controlled via the physico-chemical parameters of the solution [22-24].

**Mechanical properties of silica-latex nanocomposites**

If one wished to study the reinforcing properties of aggregates embedded in a matrix, a both simple and powerful testing procedure has to be chosen. A popular choice, especially for melts, is mechanical spectroscopy, but for solid samples uniaxial stretching is often preferred. In our case, it has the double advantage of being well suited for small angle scattering techniques, and of allowing high deformations up to rupture, forcing the nanocomposite structure to evolve. The mechanical properties of silica-latex nanocomposites have been tested by uniaxial stress-strain isotherms. The silica nanoparticles (two sizes, B30, $R_o$ = 7.7 nm, $\sigma$ = 0.186, or B40, $R_o$ = 9.3 nm, $\sigma$ = 0.279 [25]) were incorporated in a matrix containing poly(methyl methacrylate) and poly(butyl acrylate) in such proportions that the glass transition temperature was set to 33°C. The initial radius of the latex particles was $R_o$ = 13.9 nm, with a polydispersity of $\sigma$ = 0.243.

The reference case for the mechanical tests was given by the stress-strain isotherms of the pure latex films. These were found to depend strongly on pH [26], which is why the reinforcement factor presentation was chosen to highlight the filler contribution to the stress. This presentation consists in normalising at each deformation the nanocomposite stress by the stress of the pure latex sample at same pH, $\sigma(\lambda)/\sigma_{latex}(\lambda)$. It thus answers the question, how much stronger the composite with respect to its own matrix is.

In Figure 2a, the stress-strain isotherms measured with a series of nanocomposites (silica B30, pH 9, $\Phi_{si}$ = 2.5%-15%) are plotted. A strong increase in stress (and namely of the stress at small deformation) with silica volume fraction is found, whereas the ultimate properties like the extension at sample rupture are hardly changed. When choosing the reinforcement factor representation, plotted in Figure 2b for the same data, the increase in Young's modulus E, i.e. in the small deformation range, is emphasised. Young's modulus is found to rise quickly with volume fraction, up to a factor of ten for this data set. Although we will encounter even more impressive effects, this value should be compared to the much lower prediction of the Einstein/Smallwood formula [12]; $E/E_{latex} = 1 + 2.5 \, \Phi_{si} + \ldots$ . Of course this equation does not apply to high volume fractions, but the comparison illustrates the strength of the effect.

In order to quantify the increase in the low-deformation increase in reinforcement with filler volume fraction, the reduced Young's modulus $E/E_{latex}$ is shown in Figure 3. For the bigger silica beads, B40, the three data sets corresponding to different pH values are shown (pH 5, 7.5, 9) [27]. At each pH, the reinforcement factor raises quickly with $\Phi_{si}$. The strongest increase is observed at lowest pH, up to about a reinforcement factor of 40. Note that the characteristic pH-dependence of Young's modulus is also observed with the smaller silica particles, B30, shown in the inset [28]. Here the silica volume fraction is kept fixed to 5% and the pH is varied. The solid lines in Figure 3 have been calculated with a very simple theoretical expression [27]. The minimal assumptions to reproduce the data are that silica particles are aggregated, with an effective filler volume fraction described by a compacity[2] of aggregates between 10% and about 35%, and that the percolation threshold of aggregates is approximately 60%. In order to fit the data, the compacity needs to be lower at lower volume fractions and lower pH. The physical interpretation is that at low compacity there is an important amount of polymer inside the aggregates, they are therefore swollen and the hard phase approaches the percolation threshold where the modulus diverges. However, it is intuitively clear that in order to reach a weak compacity, the aggregation number must be rather big; after all, single particles have a compacity of 100%. It is thus guessed from the mechanical measurements that there must be a strong variation of the aggregation number with pH, the system forming bigger aggregates at lower solution pH.

Let us come back a moment to Figure 2b. At higher deformation, the reduced stress $\sigma(\lambda)/\sigma_{latex}(\lambda)$ seems to tend towards a plateau value which is much lower than the maximum value. This illustrates that interfacial effects are of second order. It is argued that at large deformations only hydrodynamic contributions persist, which includes the contribution of the immobilized polymer layer on the filler surface, whereas the aggregate structure is believed to be responsible of the low-deformation behaviour only. In order to check this idea we have performed structural characterization of our silica-latex nanocomposites.

---

[2] The compacity of an aggregate is defined as the ratio of the filler volume to the total aggregate volume.

**Structure of silica particles in polymer matrix**

The degree of dispersion of hard particles in a soft polymeric matrix can be determined by various techniques. In direct space, transmission electron microscopy of thin slices gives a two dimensional projection [29], whereas for example Atomic Force Microscopy can be used to image the surface [30]. One of the advantages of scattering techniques is that an average description of the structure in the bulk volume of the sample is obtained. This is why we have performed Small Angle Neutron Scattering experiments at Laboratoire Léon Brillouin in Saclay, and Institut Laue Langevin in Grenoble. The interpretation of the measured intensity in reciprocal space is not always straightforward, but nanocomposite studies using both electron microscopy and scattering techniques demonstrate that results are consistent [31]. In certain cases, like aggregation of colloidal particles, scattering can directly provide valuable information. In our case, e.g., aggregates repel each other, and their ordering within the matrix gives rise to a peak in the scattered intensity at wave vector $q_o$. This can be used to estimate the average aggregation number $N_{agg}$:

$$N_{agg} = \Phi_{si} (2\pi/q_o)^3 / V_{si} \qquad (1)$$

where $\Phi_{si}$ denotes the silica volume fraction and $V_{si}$ the average volume of a silica bead. An illustration is plotted in Figure 4, where the intensity is presented as a function of wave vector for samples with identical silica volume fraction (bigger silica beads B40, $\Phi_{si}$ = 9%) and different solution pH (5, 7.5, 9). The structure is clearly seen to depend on the pH: Using eq.(1) , the aggregation number is found to increase from 1 to 16 and finally to about 200, with decreasing pH. Although it is clear that silica particles carry less electrostatic charge at lower pH, the exact mechanism of aggregation in this rather crowded and strongly interacting solution is still unclear. Plausible mechanisms could be aggregate nucleation and growth in solution, possibly cluster-cluster aggregation, the aggregates being then imprisoned in the polymer matrix. Given the rather low filler volume fractions and evaporation temperatures, we believe that the filler structure is frozen in the matrix. In this context, an immediate experimental questions is how aggregation depends on the silica volume fraction. We have therefore measured nanocomposite structure for samples along different pH and volume fraction lines in parameter space. An example is shown in Figure 5, where the intensity curves of nanocomposites made from solutions pH fixed to 5 and varying silica volume fraction are plotted. In the log-log presentation, the intensity seems to follow approximately the same scattering law for different $\Phi_{si}$, with only a slight shift of the peak position. This indicates directly that the structure is globally similar. Using equation (1), this shift can be shown to be partly due to the change in $\Phi_{si}$, and partly to a moderate increase in aggregation number $N_{agg}$ with volume fraction, from about 200 to 300. This variation illustrates that the order of magnitude of $N_{agg}$ is set by the pH, whereas the aggregation is only weakly influenced by the volume fraction.

In order to check the influence of the silica particle size, the same structure determination has been carried out with the smaller silica beads (B30). The aggregation number of clusters made of these particles is found to have the same dependence on pH and volume fraction, however with somewhat higher aggregation numbers. To summarize this discussion of the structure of the silica particles in the soft matrix, the aggregation number for the smaller beads is plotted in Figure 6 as a function of pH and volume fraction. The result resembles a master curve, where the aggregation number is found to depend only on the solution pH. Note that a detailed examination, however, would reveal some differences in the moderate and high pH region. This does not weaken the main conclusion, which is that the solution pH has a dominating influence on the structure of the nanocomposite, capable of triggering changes of over two orders of magnitude in $N_{agg}$.

Given its importance in the analysis of the nanocomposite structure, the use of the rather simple eq. (1) for the estimation of the aggregation number has to be questioned. In the past, we have checked its validity by applying it successfully to solution structure of pure colloids with electrostatic interaction [27], where (quasi-)analytical structure factors are available [32,33]. Another, independent indication of its validity is given by the increase of the peak intensity with aggregation, because bigger aggregates scatter more [25,27]. Up to now, however, the proof that aggregates with $N_{agg}$ determined with eq.(1) do indeed scatter as experimentally observed is still missing. We will now close this gap.

The technical difficulty with the analysis of the intensities scattered from nanocomposite samples is that there are several levels of organization, two of which are unknown. At the smallest scale, the silica particle. Luckily, its shape can be measured independently and it is found to be rather monodisperse and spherical, which makes calculations easier. The silica particles are then agglomerated at a scale apparently determined by the pH. The largest accessible structure is the inter-aggregate structure factor, which translates the centre-of-mass correlations between aggregates. From the peak intensity and peak position an estimation of the aggregation number can be obtained, but unfortunately no indication on how the beads are positioned in space. In order to disentangle the different contributions, we have implemented a self-consistent structure determination based on a Reverse Monte Carlo algorithm [34, 35]. As a first important result, we have checked in several cases that an aggregate whose scattering is consistent with the experimental intensity can be constructed using the estimation for the aggregation number given by eq.(1). This aggregate is thought to represent the average structure of all aggregates in the sample. An example is shown in Figure 7, where the experimental and the calculated intensities are superimposed. Minor deviations can still be found, but we think that these could be reduced by either averaging over different aggregate structures, or by optimizing the search algorithm, or by including the resolution function of the Small Angle beamline. In any case the main features of the scattered intensity are reproduced over many orders of magnitude in intensity, which is in itself sufficient to validate our method used to estimate the aggregation numbers. Moreover, large deviations from the estimated $N_{agg}$ (more than 10 or 20%) do not lead to self-consistent solutions. Finally, a three-dimensional picture of the average aggregate is obtained. For illustration, a typical outcome of the search algorithm in direct space is also depicted in Figure 7.

We have now the first indications to answer one of the key questions in the reinforcement of theses soft/hard nanocomposites. What is the relationship between the structure and the mechanical properties of the silica-latex samples examined here ? Indeed, comparing the pH dependence of the reinforcement factor of Young's modulus (Figure 3) and the one of the average aggregation number (Figure 6) leads to an simple observation: elastomers containing big aggregates at rest show a strong reinforcement in the small deformation range, i.e. where the structure is perturbed only weakly. This is in line with the assumptions made in the mechanical results section. There we have deduced from the low deformation reinforcement that at low pH aggregates are less compact. It has been hypothesized that this can be reached only with high aggregation numbers, and we have now the confirmation that these lead indeed to higher reinforcement.

In principle, one can even go a bit further, and compare the average aggregate compacity needed to reproduce the mechanical results (solid lines in Figure 3) to the compacity of the average aggregate whose scattering is compatible with the measured intensity, cf. Figure 7.

Here the crucial question of how the latter is to be defined arises, but a quick estimation shows that the order of magnitude is indeed the same.

The next obvious question is what happens with the structure as the samples are stretched. From the mechanical measurements, the reinforcement factor (Figure 2b) is found to decrease towards a plateau value, most of the reinforcement being lost in moderate strains, below $\lambda = 2$. We have shown in a specific case that the same sample stretched a second time after a first elongation leading to rupture shows almost no reinforcement [26]. This suggests that the initial aggregate structure is destroyed, and probably replaced by more compact aggregates. Up to now, only very few measurements of the structure under deformation have been published [22,36]. The main reason is that the interpretation of the data is rather difficult. An example of a two dimensional intensity map is shown in Figure 8. On the left, the isotropic intensity measured from a sample of silica B40, at pH 5 and $\Phi_{si} = 6\%$, at rest is shown. Its radial average leads to the low-q part of the corresponding function plotted in Figure 5 (several sample-to-detector distances are necessary to construct one intensity function), and the aggregation number is determined to be about 170. The sample has then been stretched to $\lambda = 1.7$ at 60°C, i.e. above its glass transition, and immediately been cooled below it. It is placed vertically in the neutron beam, and the scattered intensity is shown on the right-hand-side of Figure 8. The intensity looks like a ellipse whose major axis is perpendicular to the drawing direction. From the ratio of its axes the microscopic stretch ratio, on the scale of a few hundred nanometers, can be estimated to be approximately 1.6, i.e. close to but smaller than the macroscopic elongation. In the literature, this is termed 'non affinity' of the microscopic motion with the macroscopic strain. The signature of affinity would be an ellipse deforming at the same strain as the sample. It is not clear, however, what exactly causes this non affinity in the present case. It is possible that aggregates break up or deform less than the total sample, but probably they also collide and reorganize at the scale of the inter-aggregate structure factor. Other, more exotic signatures of non affinity have been observed, like butterfly patterns and four-spot intensity maps [22]. In analogy with intensity maps of rubber with spatial heterogeneities in the degree of crosslinking, butterfly patterns are generally explained as being due to harder areas within the sample [37]. These resist deformation, which leads to larger strain of softer areas, which is in turn thought to be related to the reinforcement effect [5]. From a structural point of view, butterfly patterns thus seem to be related to aggregation. A measurement in favor of this interpretation is plotted in Figure 9.

The intensity maps at rest and at $\lambda = 2.0$ of a sample containing the smaller silica beads, B30, at 10% volume fraction, and low pH 5.2 are compared. The strong ordering can be seen in the strong peak yielding a ring on the intensity map, and the aggregation number at rest is of the order of 320 [3]. Upon deformation, shown on the right-hand side of Figure 9, the intensity weakens considerably in the perpendicular direction, whereas two prominent butterfly wings above and below the beamstop are easily recognized. Unfortunately the structures shown in Figure 8 and 9 have been measured with samples which are too different for a direct comparison. This example suggests, however, that big aggregates move independently in the matrix at lower volume fraction (and deform somewhat less than the macroscopic strain), whereas big aggregates in the more concentrated sample are closer to collision and percolation. In the former case, the structure factor deforms almost in an affine manner, yielding an ellipse in the intensity map, and in the later case the famous butterfly patterns are recorded.

It has been attempted to reproduce anisotropic spectra using laws of motion for filler particles dispersed according to certain distributions [38]. Main features and effects visible in two dimensional spectra, like departure from affine deformation, have been accounted for by simple collision laws. A typical result is shown in Figure 10, where the structure factor of a two dimensional assembly of beads submitted to uniaxial strain is plotted. A four-spot intensity map is found, qualitatively similar to experimental intensities as presented in ref. [22]. Quantitative analysis, however, was not feasible using this approach, mainly due to the complexity in describing aggregate collisions. A new and completely different method might be the extension of the Reverse Monte Carlo algorithm to two dimensional intensity maps. Progress in this field will allow to address open issues like the degree of affinity of the motion of aggregates, and the interplay between aggregate motion and deformation.

---

[3] A first (too high) estimation has been reported in ref. [26], whereas the present measurement has been performed with a more suitable beamline for very small angle scattering.

**Conclusion**

We have reviewed recent results on the structure of aggregates of nanometric silica spheres embedded in a polymer matrix using the latex route. The average aggregation number is found to be tuneable via the precursor solution pH, and to depend only marginally on the total silica volume fraction. These aggregates have been shown to have a strong influence on the mechanical properties of the nanocomposite. Bigger aggregates show a stronger reinforcement at constant hard filler volume fraction. Our interpretation is that these aggregates are less compact, which is why they bring the system closer to the percolation threshold of hard material. This leads to a strong reinforcement of Young's modulus, as hard domains can collide, reorganize, and maybe percolate. In this context, the possible existence of an immobilized polymeric layer at the surface of the particles needs to be discussed [6-8]. We think that the influence of this interfacial layer would be of second order with respect to the structural contribution because the aggregates span (and partially immobilize) a much larger volume of polymer than the surface layer on peripheral aggregate particles. At large strains, however, the immobilized layer increases the hydrodynamic contribution of each particle.

Although we have studied the structure – mechanical properties relationship of a soft/hard nanocomposite showing a strong reinforcement effect in some detail, many open questions remain. As an example, the process of the aggregate formation in a crowded environment remain unknown. Given the difficulties in following the aggregation in the drying solution, it is hoped that numerical simulations will contribute to the understanding of the time-dependent structure of the strongly interacting solutions in the future.

Another mystery is given by the modes of deformation of the aggregates inside a stretched sample. In scattering experiments, clear evidence for non affine displacement of the filler particles is obtained, but in spite of some progress using computer simulations, interpretation is still difficult. Maybe the solution will come from Reverse Monte Carlo algorithms, or from observations in direct space, e.g. by AFM on the sample surface [39].

To conclude it is recalled that one important property of the silica-latex system is that the interface is almost unchanged, which allows us to relate aggregation directly to mechanical reinforcement. A controlled modification of the interface by grafting polymer chains opens then a new field of study. Indeed, such a modification is expected to influence both the aggregation behaviour in polymer matrices and the filler-matrix interaction. Many groups have therefore used polymer to link the filler to the matrix, e.g. ref. [6-8]. Recent advances in controlling the interfacial properties of the filler particles at the level of the molecule have been made using a controlled radical polymerisation technique (Atom Transfer Radical Polymerisation), to graft poly(styrene) and poly(butyl methacrylate) chains from the silica surface [40-42]. At the moment, the synthesis has been set up in organic solvents, which is why the grafted filler particles were chosen to be incorporated directly in a hydrophobic matrix polymer [43]. A scope for the future, however, would be the use of water soluble chain molecules in order to use the same latex route of preparation of nanocomposite samples the success of which we have briefly outlined in this article.

**Acknowledgements :** It is a pleasure to acknowledge support and encouragement by the Small Angle Staff of LLB, collaboration with François Boué, and help by Bruno Demé at ILL. Parts of this work were conducted within the scientific programme of the Network of Excellence 'Soft Matter Composites: an approach to nanoscale functional materials' supported by the European Commission. Fruitful discussions with Peter Hine (Leeds) and Wim Pyckhout-Hintzen (Juelich) in the framework of this network are acknowledged. Silica and latex stock solutions were a gift from Akzo Nobel and Rhodia.

**Figure Captions :**

Figure 1 :    Drawing illustrating the latex route for incorporation of nanoscopic colloidal silica bead in polymer films by latex film formation.

Figure 2 :    (a) Stress-strain isotherm $\sigma(\lambda)$ of silica-latex nanocomposite (silica B30, pH 9, $\Phi_{si}$ = 2.5%-15%). (b) Reinforcement factor $\sigma(\lambda)/\sigma_{latex}(\lambda)$ of the same series. (Reprinted with permission from ref. [26], copyright 2002, American Chemical Society).

Figure 3 :    Small-deformation reinforcement factor $E/E_{latex}$ as a function of silica volume fraction, for different solution pH (silica B40, pH 5 ($\circ$), pH 7.5 ($\bullet$), pH 9 ($\square$)). The solid lines are model calculations, see ref. [27] for details. (Reprinted with permission from ref. [27], copyright 2005, Elsevier). In the inset, $E/E_{latex}$ is shown as a function of pH, for a different silica bead (silica B30, $\Phi_{si}$ = 5% ). (Reprinted with permission from ref. [28], copyright 2004, Springer).

Figure 4 :    Structure of nanocomposites as seen by Small Angle Neutron Scattering. Scattered intensity I(q) is plotted for three samples at identical volume fraction, different precursor solution pH (silica B40, $\Phi_{si}$ = 9%, pH = 5, 7.5, 9). Arrows indicate the peak position. (Reprinted with permission from ref. [28], copyright 2004, Springer).

Figure 5 :    Scattered intensity I(q) from nanocomposites of silica B40 at fixed pH = 5.0, for increasing volume fraction $\Phi$. In the inset, the low-q intensity is shown in linear scale in order to emphasize the strong ordering. (Reprinted with permission from ref. [25], copyright 2002, American Chemical Society).

Figure 6:     Average aggregation number of aggregates of B30 silica beads in nanocomposite samples estimated from the position of the intensity peak, as a function of solution pH. Results at different silica concentrations are superimposed (silica B30$\Phi_{si}$= 5%, 10%, 15%). (Reprinted with permission from ref. [27] , copyright 2005, Elsevier)

Figure 7:    Structure of a nanocomposite (B30, pH 7, $\Phi_{si}$= 5%). The experimental intensity I(q) is compared to a model prediction based on a Reverse Monte Carlo algorithm with inter- and intra-aggregate structure factor. The aggregate structure plotted in the inset illustrates the typical solution. The aggregate is continuous, space between particles is due to the polydispersity which is not reproduced by the programme used for the drawing.

Figure 8 :    Intensity map of a nanocomposite sample (silica B40, pH 5, 6%, cf. Fig. 5) measured at wavelength 10 Å, detector distance 36.7 m. (a) The isotropic sample, (b) The same sample stretched vertically to a draw ratio of $\lambda = 1.7$.

Figure 9:    Intensity map of a nanocomposite sample (silica B30, pH 5.2, 10%) measured at wavelength 10 Å, detector distance 36.7 m. (a) The isotropic sample, (b) The same sample stretched vertically to a draw ratio of $\lambda = 2.0$.

Figure 10 :    Anisotropic two dimensional total structure factor at elongation ratio $\lambda = 2.5$, showing four maxima in the intensity. It is obtained from simulations of mutually avoiding spheres through local shear. $\Phi_{surf}$ =15%, polydispersity 22.2%, R = 90 Å. Axes are in Å $^{-1}$. (Reprinted with permission from ref. [38], copyright 2000, Elsevier).

**Figures**

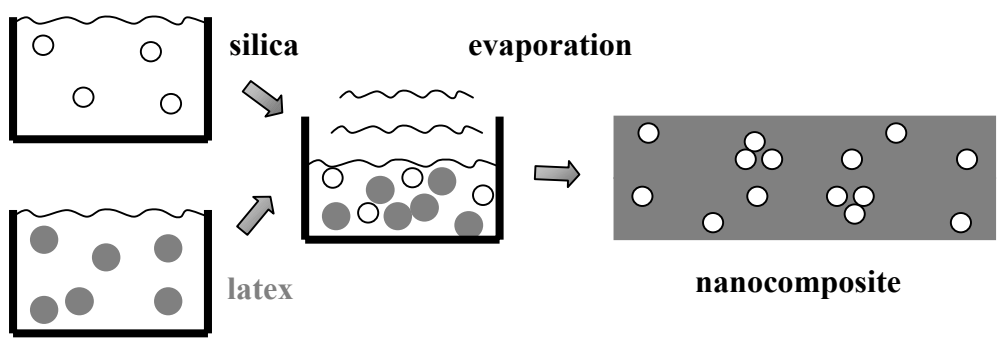

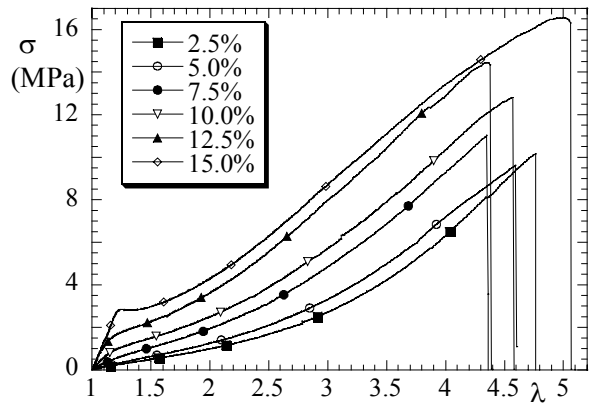 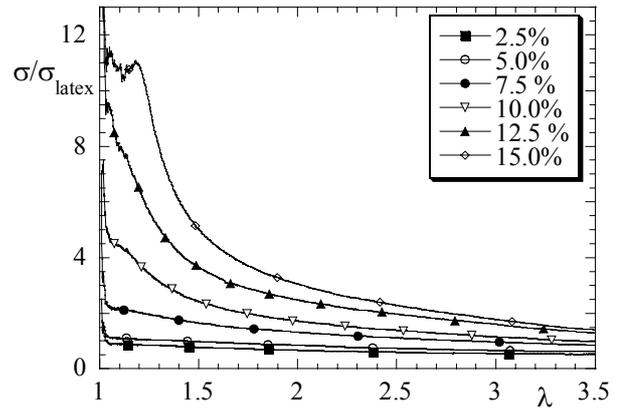

(a)                                      (b)

FIGURE 2 (OBERDISSE)

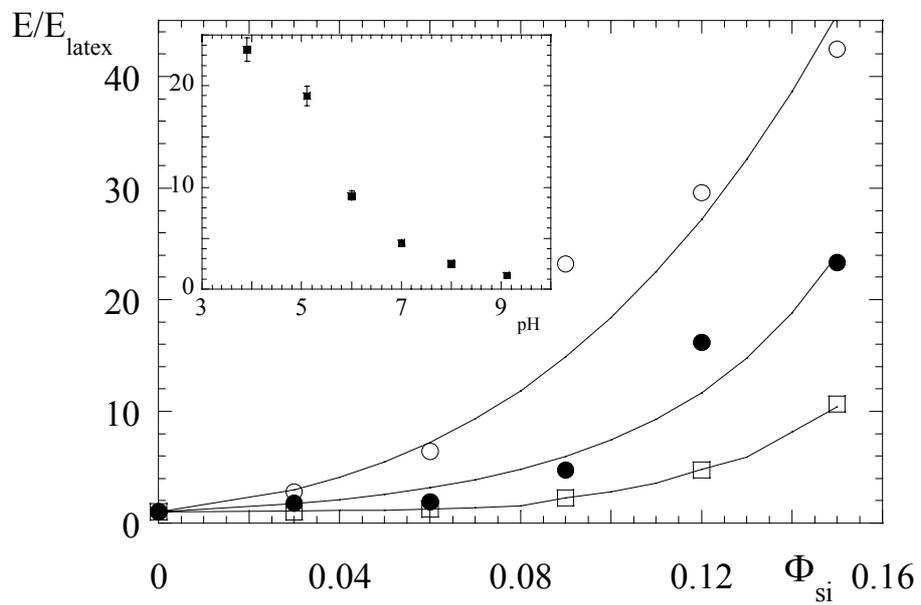

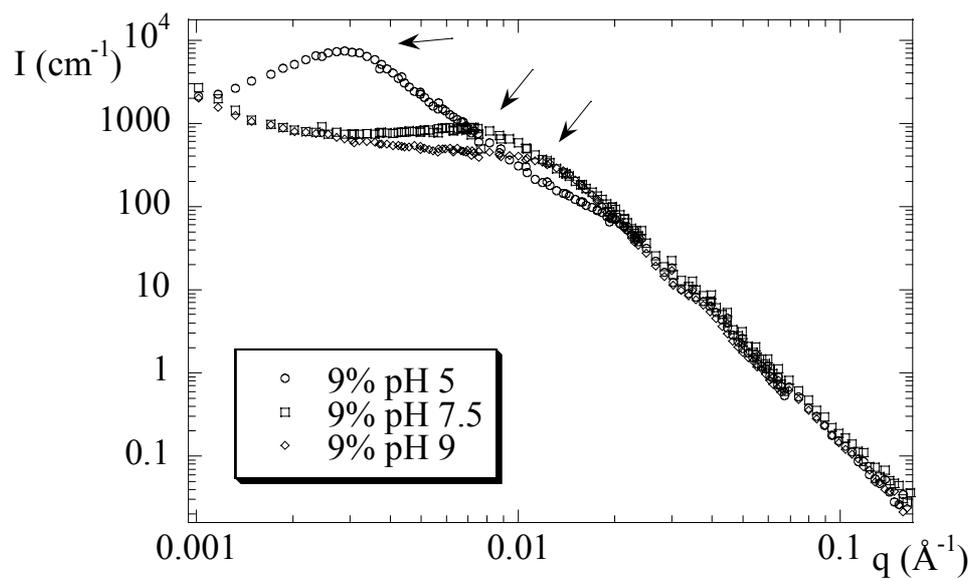

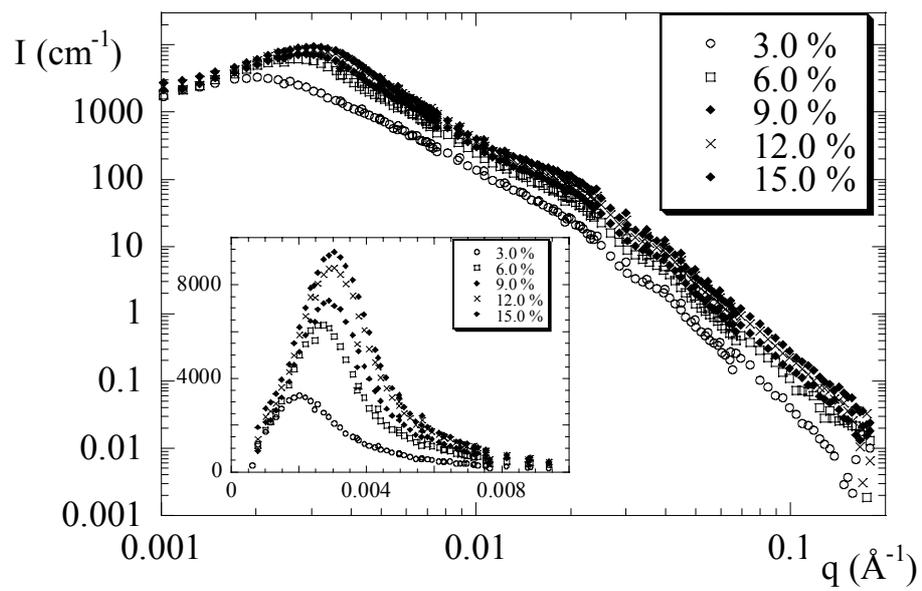

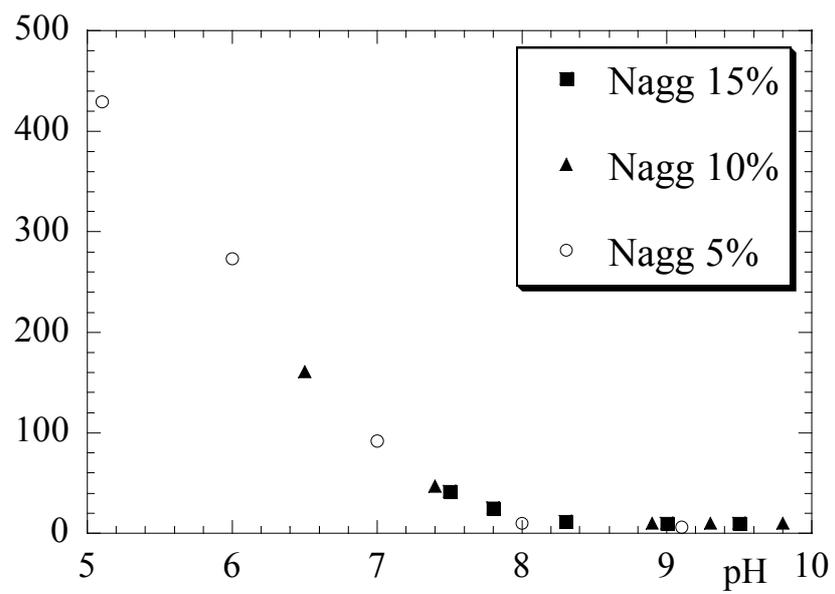

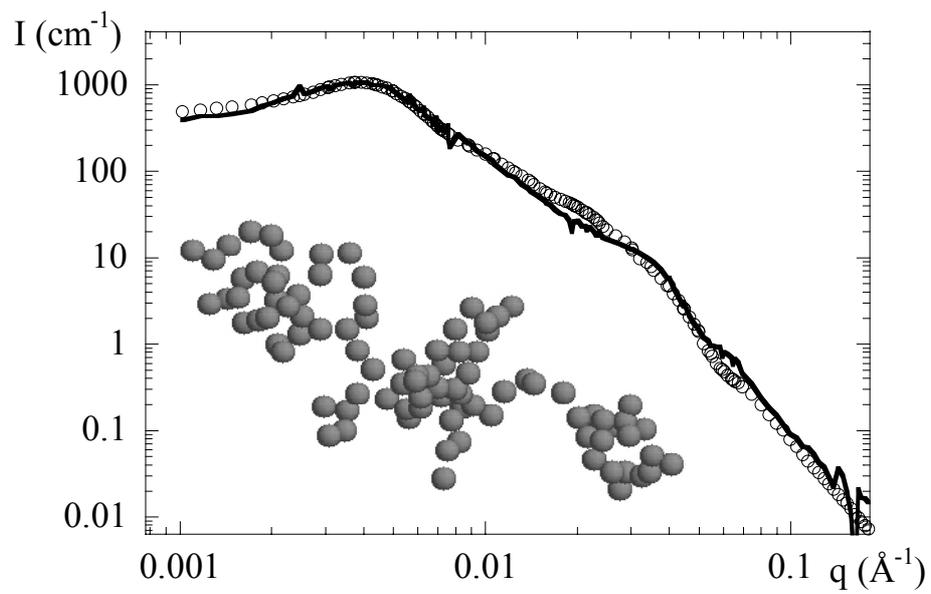

I (cm⁻¹)

q (Å⁻¹)

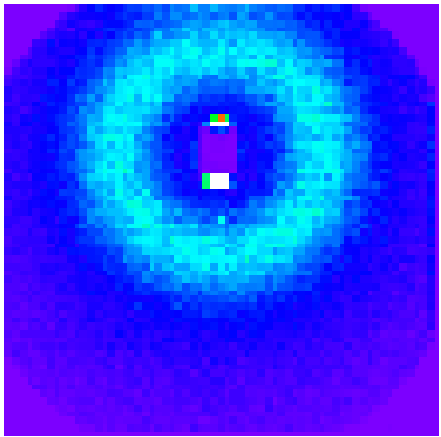 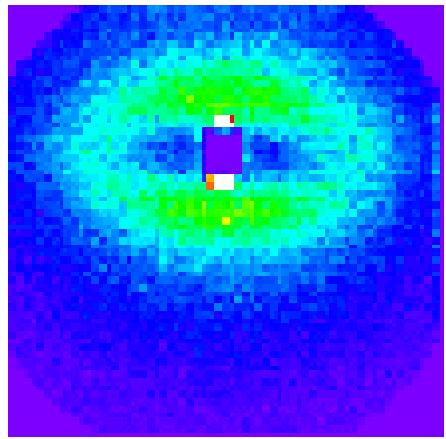

(a)                                              (b)

FIGURE 8 (OBERDISSE)

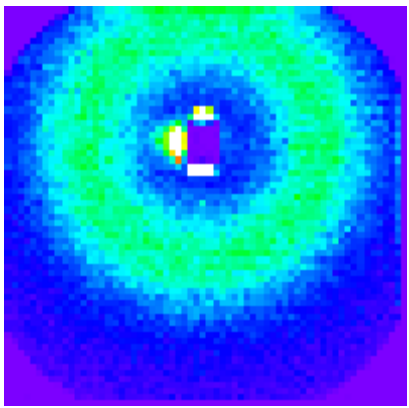 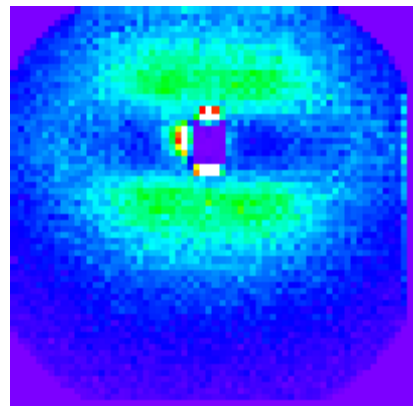

(a)                                  (b)

FIGURE 9 (OBERDISSE)

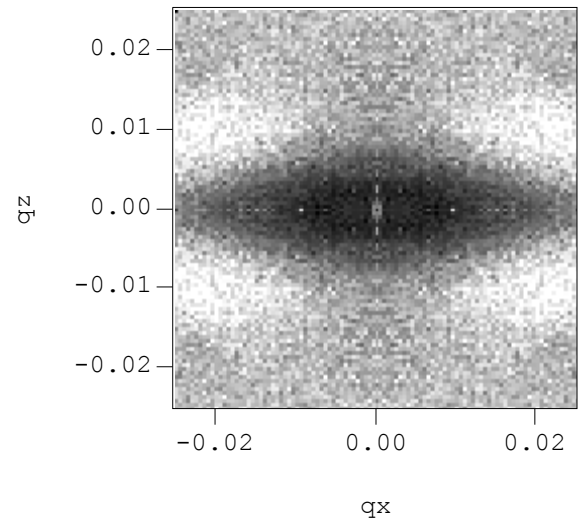

FIGURE 10 (OBERDISSE)